\documentclass[prl,preprint,showpacs,epsf]{revtex4}
\usepackage{graphicx}
\usepackage{color}
\begin{document}
\draft
\title
{A New  Precision Measurement of the $^3$He($^4$He,$\gamma$)$^7$Be Cross 
section}
\author{
B.S.~Nara Singh$^1$, M.~Hass$^1$, Y.~Nir-El$^2$ and G.~Haquin$^2$}
\affiliation{
{ 1.~~Department of Particle Physics, Weizmann Institute of Science, 
Rehovot, Israel}\\
{ 2.~~Soreq Research Centre, Yavne, Israel}\\
}
\date{\today}

\begin{abstract}

{
  The $^3$He($^4$He,$\gamma$)$^7$Be reaction plays an important role
  in determining the high energy solar neutrino flux and in
  understanding
  the abundances of primordial $^7$Li.
The present paper reports 
a new precision measurement of the cross sections of this 
direct capture reaction, determined by measuring the 
ensuing $^7$Be activity in the region of $E_{c.m.}$~=~400~keV 
to 950~keV.  Various recent theoretical fits to our data 
result in a consistent extrapolated value of 
$S_{34}(0)$~=~0.53(2)(1).


}
\end{abstract}
\maketitle

The $^3$He($^4$He,$\gamma$)$^7$Be reaction is one of the 
remaining major sources of uncertainty~\cite{Bah1, Bah2} in 
determining the high energy solar neutrino flux~\cite{SK,SNO} 
that results from the $^7$Be(p,$\gamma$)$^8$B reaction
\cite{WIS,Seat,CD1}.  It also plays an important role in 
understanding the primordial $^7$Li abundance~\cite{Nol,Cybu}.
The available data on the astrophysical $S$-factor $S_{34}$ 
are obtained by using two different methods.  Namely, the 
detection of prompt $\gamma$-rays~\cite{Hilg, Park, Naga, Kraw, Osbo, Alex}
from $^{7}$Be or of the ensuing  $\gamma$-activity from ~$^7$Be
~\cite{Hilg, Osbo, Robe, Volk}.  These two sets of results show 
a significant scatter and a persistent discrepancy~\cite{Adel},  
resulting in the presently recommended low-precision values
~\cite{Adel,Angu}.  The Standard Solar Model (SSM) 
calculations~\cite{Bah1, Bah2} and Standard Big Bang 
Nucleosynthesis (SBBN)~\cite{Coc} use 0.53(5)~keV-b 
of Ref.~\cite{Adel} and 0.54(9)~keV-b of Ref.~\cite{Angu},
respectively, for $S_{34}$(0). The most recent compliation
~\cite{Desc} quotes a value of 0.51(4)~keV-b. A more accurate measurement 
is therefore highly desirable and recommended~\cite{Bah2}.
However, almost no such attempt was made for almost two 
decades, even though the $^7$Be(p,$\gamma$) reaction
has drawn much effort in the last several years~\cite{WIS,Seat,CD1}. 
Since the previous activity measurements were carried out at only one
or two energies each, we have  initiated a $^7$Be activity precision
measurement of this cross section at energies around
$E_{c.m.}$~=~400~-~950~keV using a $^{3}$He beam and a gas target
with a Ni-foil window.
Even though the method of using a gas foil is
manifestedly not suitable for determining $S(E)$ at very low energies, 
the focus of the present measurement is to obtain  accurate
data points at medium energies in order to set the absolute
scale of the cross section and for a comparison to previous measurements.

\begin{figure}[t]
\hspace{-14mm}
\includegraphics[width=9cm]{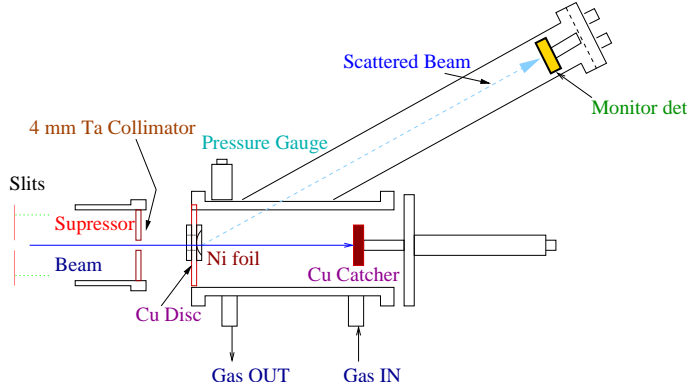}
\hfill \caption{ A schematic diagram of the experimental setup.}
\label{expsetup}
\end{figure}

\begin{figure}[t]
\includegraphics[width=9cm]{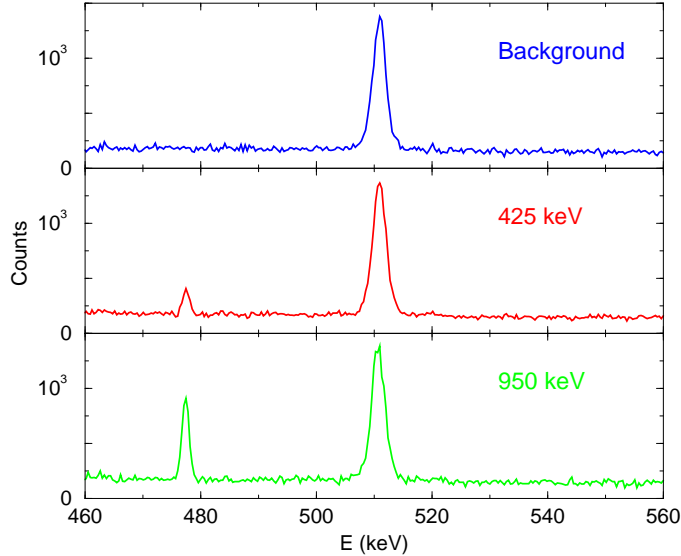}
\hfill \caption{ $\gamma$ spectra from ambient background and from the
  Cu catchers of $E_{c.m}$~=~425 and 950~keV.
The spectra are normalized to the background 511~keV, 
the only visible $\gamma$ line in this energy range besides the
478~keV from $^7$Be.}
\label{7be_act}
\end{figure}

A schematic diagram of our experimental setup is shown in 
Fig.~\ref{expsetup}.  A $^3$He beam at $E_{lab}$~=~1000 to 
2300~keV from the 3 MV Van de Graaff accelerator at the Weizmann 
Institute enters a $^4$He gas cell through a 8~$mm$ diameter 
nickel window typically of 0.5 to 1~$\mu m$ thickness.  The beam 
direction is defined by an upstream slit at 2 meters from the 
center of the cell and two Ta collimators of 3~$mm$, one at the 
entrance and the other at the exit of the chamber.  The beam on 
target is restricted to be below 1~$\mu A$ current and is 
raster-scanned over a rectangular area of 3~$\times$~5~mm in 
order to avoid excessive localized heating of the Ni window. 
The gas cell is insulated from the beam line and the 
entire chamber, including a Cu catcher of 50 mm diameter that is in 
electric contact with the gas cell, serves as a Faraday 
cup.  An aperture placed before the Ni window, including a 
4~$mm$ Ta collimator, serves as a secondary electron 
suppressor (Fig.~\ref{expsetup}) at -400~$V$ that was set by 
measuring the beam current on the chamber as a function of 
voltage; there was no discernable variation in beam current
upon introducing gas into the chamber. The elastically scattered 
beam particles from the Ni window were monitored on-line using a 
narrowly-collimated Si surface barrier detector placed at an angle of 
$\theta$~=~44.7$^{\circ}$.  The Ta collimators constrain the 
beam direction to coincide with the chamber-axis and thus 
determine $\theta$.  The beam energy was calibrated using the 
$^{27}$Al(p,$\gamma$)$^{28}$Si resonances at proton energies 
of 1118.4, 991.2 and 773.7~keV using H$_2$ and H$_3$ beams, 
respectively, in order to correspond to the range of $^{3}$He lab 
energies and magnetic rigidity.  The  
energy losses and energy straggeling of the $^3$He beam in the
Ni foil ($\Delta E_{Ni}$) 
and in the $^4$He gas ($\Delta E_{He}$) were determined using 
TRIM~\cite{Trim} and were also checked using the 1.518 MeV resonance 
in the $^{10}B(\alpha,p)^{13}C$ reaction (see below).  The center 
of mass energy for $^3$He at the center of the $^4$He gas is given by,
\begin{equation}
E_{c.m.}~=~\frac{4}{7}~(E_b~-~\Delta E_{Ni}~-~\frac{\Delta E_{He}}{2})
\label{Ecm}
\end{equation}
where $E_b$ is the beam energy.  The $^7$Be fusion product nuclei move 
forward in the laboratory system and are implanted in the Cu catcher 
at depths of few microns.  In the present energy range, the back 
scattering loss of these implanted nuclei is not relevant~\cite{Weiss}.  
The catcher distance from the Ni foil and the $^4$He gas pressure 
inside the cell are accordingly adjusted to obtain optimum 
target thickness for the measurement at a given energy.  For 
example, at $E_{c.m.}$~=~950 keV, the energy width of the target, 
$\Delta E_T~=~\frac{4}{7}~\Delta E_{He}~ \sim$~100~keV.  For all beam 
energies, down to the lowest energy of $E_{c.m.}$~=~420 keV, these 
conditions ensured that the measured cross section, integrated over 
the energy range $E_{c.m.}~-~\Delta E_T/2~<~E~<~E_{c.m.}~+~\Delta E_T/2$, 
is a true representation of $\sigma(E_{c.m.})$ at the center of 
the energy range to better than 2~$\%$ (see below).  The target gas of 
99.9$\%$ purity was monitored and maintained at a constant pressure.
The number of $^4$He target atoms per $cm^2$ is given by, 
$N_t~=~9.66~\times~10^{18}~~l~~\frac{P}{T_0~+T_c}$ where $P,~T_0,~T_c$ 
and $l$ are gas pressure, room temperature, correction in temperature 
due to the beam heating and target length given in units of $torr$, 
$^\circ  K$ and $cm$, respectively~\cite{Osbo, Gorr}.  For the typical 
500~$n$A current at 2~MeV beam we have estimated a value of 
$\approx~17~^\circ$K for $T_c$ that was confirmed using the 
1.518~MeV resonance in the $^{10}B(\alpha,p)^{13}C$ reaction.  
$^7$Be decays to the 478~keV state in $^7$Li with 
$T_{1/2}$~=~53.29~(7)~days~\cite{Tul} and a branching ratio of 
10.52~(6)~$\%$~\cite{Chu}.  The number of produced $^7$Be nuclei is 
determined by measuring the $^7$Be activity on a Cu catcher at Soreq
Research Centre using a Ge detector setup, similar to that used in the 
precision determination of the $^7$Be target strength for the $S_{17}$ 
measurement~\cite{WIS}.  To cover a large solid angle the Cu catcher 
was placed at a distance of 20~$mm$ from a  HPGe detector that was 
shielded from room background; the activity was measured  over a 
period of 3 to 6 days.  A $^7$Be reference point source of 2~$mm$ 
diameter with a precisely known $\gamma$-ray emission rate~\cite{WIS} 
was used for the efficiency calibration at 478~keV.  The $^7$Be
products subscribe approximately  a 20~$mm$ diameter spot
(TRIM~\cite{Trim}), still much smaller than the catcher's diameter,
and a correction factor of 1.3~$\%$ to the 
detection efficiency due to extended source had to be determined by 
measuring the count-rate of the reference point source at off-center 
locations. The resulting overall detection efficiency at this geometry 
was determined to be 0.0436(5) at 478~keV.  The ambient background was 
monitored periodically with a sample of 200 $cm^3$ triple-distilled 
water and a typical spectrum accumulated over $\sim$~4~days is shown 
in Fig.~\ref{7be_act}, demonstrating that there is no interfering peak 
around 478~keV.  $\gamma$-spectra measured with Cu catchers prepared at 
$E_{c.m.}$~=~425 and 950~keV are shown in Fig.~\ref{7be_act}.  The 
number of $^7$Be nuclei $N_{{^7}Be}(0)$ at the time of production is 
a measure of the cross section $\sigma$ and is obtained from the 
efficiency corrected 478~keV $\gamma$ yield and the known branching 
ratio and half life.  
For example, for the low-statistics spectrum corresponding to  
$E_{c.m.}$~=~425~keV, the net peak area, the activity and number of 
$^7$Be atoms were 898~(54) counts, 0.389~(24)~Bq, and 
2.59~(16)~$\times~$10$^6$ atoms, respectively.  The total error 
includes also uncertainties of detection efficiency from the $^{7}$Be 
reference source, counting geometry and radial distribution of the 
$^7$Be on the Cu catcher. 
The number of beam particles, $N_p$ obtained either through the beam 
current integration or the Rutherford scattering from the Nickel foil
is the other major source of error in the determination of the 
cross section.  For the scattered flux measurement ($\sim$1.8~$\%$), the 
sources of error include, a)~$\theta$~($<$~0.5~$\%$) and 
d$\Omega$ (1.1~$\%$) of the Si monitor detector.  $\theta$ was 
cross-checked using the elastic scattering at different energies of 
$^4$He beam from $^{12}$C foil.  For d$\Omega$ determination, the 
diameter of the collimator used on the particle detector was 
measured using an $\alpha$ source as $\approx$~0.217~$mm$ relative to 
a mechanically well measurable reference collimator of diameter~4~$mm$. 
b)~Ni foil thickness (2~$\%$), measured by weighing, and from alpha 
particle energy losses.  These were cross-checked with width 
measurements using an electron microscope.  The errors in $E_{c.m.}$ 
result mainly from Ni foil thickness and have major contributions to 
the $S$-factors ($2~-~7~\%$).  These were determined for each measurement 
(table~\ref{Sfac}) from peak position and width of the scattered-beam spectra,
with and without gas ($E_b$), TRIM calculations ($\Delta E_{Ni}$,
$\Delta E_{He}$) (1.5~$\%$). 

Other sources of error include, 1)~gas pressure ($<$~0.5~$\%$), 2)~bowing 
effect on the gas cell length resulting from the pressure difference 
between the beam line and the chamber ($<$~0.5~$\%$), 3)~$T_c$ ($<$~1.0~$\%$) 
and 4)~Current integration ($\sim$~1.2$\%$).  The current integration 
and the scattered particles were compared continuously and were found 
to be stable within a mean deviation of 1.2~$\%$.  These errors were 
of similar magnitude for all our measurements,  yielding a total error 
of $1.7~\%$.  In table~\ref{Sfac} we present the cross sections 
obtained utilizing both current integration (1.7~$\%$), $\sigma$ as 
well as the Rutherford scattering (2.2~$\%$), $\sigma^R$.  As evident 
from table~\ref{Sfac}, the extracted  values obtained from both 
$\sigma^R$ and $\sigma$ show no major differences and the latter is used, 
with negligible consequences on the extracted  final result for S$_{34}$(0). 

The measurements were carried out at $E_{c.m.}$~=~425, 495, 605, 610 
and $\approx$~950~keV.  The latter energy point was remeasured several times
by varying experimental parameters such as target gas pressure, beam 
current and Ni foil thicknesses, yielding  consistent values 
(see Table ~\ref{Sfac}).

The astrophysical $S(E)$ factor is related to the cross section 
$\sigma (E)$ by:
\begin{equation}
S(E)~=~E~\sigma (E)~\exp{(2~\pi~\eta)}
\end{equation}
where 2~$\pi~\eta$~=~164.12/$E^{1/2}$ and $E$ is given in keV.
Table~\ref{Sfac} presents measured cross sections and extracted 
$S$-factors at various $E_{c.m.}$.  To examine the issue of the 
absolute scale of various measurements, we have carried out a 
$\chi ^2$ compatibility analysis  between $S(E)$ values from 
previous data sets (grouped together in the vicinity of the 
present energies) and the present results (Table ~\ref{Indata}). 
It is evident that only the results of Hilgemeier 
{\it et al.}~\cite{Hilg} are fully consistent with ours without the
addition of an extra re-normalization parameter.

\begin{figure}[t]
\includegraphics[width=7.7cm]{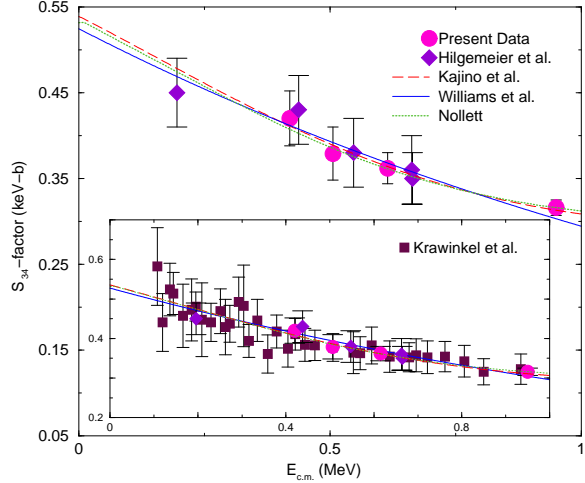}
\hfill \caption{Present data together with that of  ~\cite{Hilg} and
representative theoretical fits, yielding $S_{34}$(0)~=~0.533
(20)(7)~keV-b. Inset: the results of renormalized
Ref.~\cite{Kraw} (see text) are also included to yield  0.532(30)(4)~keV-b.}
\label{tomb}
\end{figure}

\begin {widetext}

\begin{table}[ht]
\caption{Capture cross sections at different values  of $E_{c.m.}$, Ni 
windows, target gas pressure ($P$) and length ($L$).  $\sigma ^{R}$ 
and $\sigma$: The cross sections values  obtained by using the 
number of beam particles from Rutherford scattering and charge 
integration, respectively.  $S(E)$: $S$-factors corresponding to 
$\sigma$.  The errors due to statistics (activity measurements), 
and systematics ($P$, $T_c$ and $E_{c.m}$) are also given in 
separate brackets. The latter errors on S(E) include also the
uncertainties on energy losses and straggeling (see text).} 
\begin{center}
\begin{tabular}{ccccccc}
\hline
\hline
$E_{c.m.}~(keV)$ & Ni~($\mu m$) & $P (torr)$ & $L (cm)$ 
& $\sigma ^{R}~(nb)$ & $\sigma ~(nb)$ & $S(E)~(keV-b)$\\
\hline
\hline
951.0 & 1.00 & 50.0 & 10.33 & 1680 (59) (37) & 1680 (59) (29)& 0.328 (12) (7) \\
951.0 & 1.00 & 36.8 & 13.66 & 1500 (45) (33) & 1530 (46) (26)&  0.299 (8) (7) \\
951.0 & 1.00 & 34.9 & 13.93 & 1830 (75) (40) & 1720 (71) (29)& 0.335 (14) (9) \\
951.0 & 1.00 & 52.8 & 10.35 & 1700 (76) (37) & 1600 (72) (27)& 0.312 (14) (8) \\
950.0 & 0.50 & 51.3 & 10.35 & 1580 (57) (35) & 1690 (61) (29)& 0.330 (12) (9) \\
950.0 & 1.00 & 50.4 & 10.35 & 1518 (58) (33) & 1586 (61) (27)& 0.309 (12) (8) \\
$\overline {950.0}$ & - & - & -& 1620 (31) (37) & 1620 (31) (29) & 0.316 (6) (7) \\
624.0 & 1.00 & 50.0  & 10.35 & 764 (30) (17)& 794 (31) (14)&  0.353 (14) (17) \\
605.0 & 0.50 & 50.0  & 10.35 & 767 (31) (17)& 777 (31) (13)&  0.372 (15) (15)\\
$\overline {614.5}$ & - & - & - & 766 (22) (17) & 786 (22) (13) & 0.362 (10) (15)\\
506.0 & 0.75 & 22.4  & 10.35 & 476 (16) (10)& 508 (17) (8)&  0.379 (15) (27)\\
420.0 & 1.00 & 20.4  & 10.35 & 303 (9) (7)& 333 (10) (6)&  0.420 (14) (30)\\
\hline
\end{tabular}
\end{center}
\label{Sfac}
\end{table}

\end {widetext}
\newpage
\begin{table}[ht]
\caption{A $\chi ^2$ comparison of $S$-factors from the present data and 
from former measurements, interpolated for the energies of the current 
experiment.  The $S_{34}(E)$ from Ref.~\cite{Kraw} have been scaled up by 
$40~\%$, as suggested in Ref.~\cite{Hilg}. The data from Ref.~\cite{Osbo}
include both prompt $\gamma$ and $^7$Be activity measurements.}
\begin{center}
\begin{tabular}{cccccc}
\hline
\hline
$E_{c.m.}$~(keV) & Present & ~\cite{Kraw} & ~\cite{Hilg} & ~\cite{Osbo} & ~\cite{Naga,Park}\\
\hline
\hline
420.0 &  0.420(32)& 0.38(2)& 0.44(4)  & 0.38(1)  &0.42(2) \\
506.0 &  0.379(31)& 0.36(1)& 0.40(4)  & 0.39(1)  &0.35(2) \\
615.0 &  0.362(18)& 0.34(2)& 0.36(4)  & 0.40(1)  &0.37(2) \\
950.0 &  0.316(9)& 0.30(2)& 0.28(4)  & 0.36(2)  &0.26(1) \\
$\chi^2$ &   ---  & 2.7  & 1.1  & 9.0 & 18.0 \\
\hline
\end{tabular}
\end{center}
\label{Indata}
\end{table}
The various theoretical models~\cite{Tomb, Liu, Kim, Wil, Laga, Kaji, Adel} 
are normally constrained  by nuclear model parameters that reproduce 
measured nuclear properties such as binding energies, branching
ratios, charge radii and electric quadrupole moments and largely yield a
similar energy dependence of the $S(E)$ factor.  
Due to remaining 
ambiguities, the overall normalization is left as a free parameter to be 
fitted to the data, subject to the  constraint: 
0.4 keV-b~$\leq~S(0)~\leq$~0.9~keV-b ~\cite{Kaji}. 
These fits (Fig.~\ref{tomb}) yield extrapolated values of 
S$_{34}$~=~0.53(2)(1) and S$_{34}$~=~0.53(3)(1) for the present data
alone and when combined with the  data from Ref.~\cite{Hilg}, respectively. 
The errors in brackets represent the experimental error and the 
variation of the extrapolated $S_{34}$(0) using a particular theory,
respectively.  The experimental error includes also the $\approx~2~\%$ 
uncertainty in relating $\sigma(E_{c.m})$ to the mean $\sigma$ as 
discussed above.  It is also instructive to include the extensive 
data set from Ref.~\cite{Kraw}, that exhibits a similar energy 
dependence of $S(E)$, in the fit to the present results, with the 
addition of an inter-set normalization parameter, yielding
$S_{34}$~=~0.532(30)(4).  The excellent agreement of the present 
values with the prompt-$\gamma$ values of Ref.~\cite{Hilg} and 
with the renormalized data of Ref.~\cite{Kraw} results in a
statistical agreement
between prompt-$\gamma$ and decay-$\gamma$ measurements ~\cite{Adel}.  
The procedures outlined above are all consistent and we quote a final
recommended result of  $S_{34}$(0)=0.53(2)(1) keV-b.  

The $S_{34}$(0) value used in the current SSM~\cite{Bah2} yields a 
8~$\%$ uncertainty in the predictions of both the $^7$Be and $^8$B 
neutrino fluxes.  Since these fluxes are proportional to 
$S_{34}^{0.86}$(0) and $S_{34}^{0.81}$(0), respectively~\cite{Bah3},  
the current result of $S_{34}$(0)~=~0.53(2)(1) will bring down the 
uncertainty to a level of 5~$\%$.  The quoted value of $S_{34}$(0) 
also provides a more accurate and reliable input for the SBBN 
simulations.  The present recommneded value is in excellent agreement
with the most recent updated NACRE complication of 0.51(4)~keV-b
~\cite{Desc}. We note that the agreement with ~\cite{Desc} is even more
remarkable if one uses the normaliztion to the data of ~\cite{Kraw} as
outlined above.The present value highlights the issue of the marked discrepancy 
between the calculated $^7$Li abundance using the baryon density from 
Cosmic Microwave Radiation measurements and observations~\cite{Coc}
and emphasizes the need for another resolution to this discrepancy.

We thank Y.~Shachar and the technical staff of the accelerator 
laboratory  at the Weizmann Institute for their help and support.  
We acknowledge C.~Bordeanu for her work on the initial design of 
the setup and I.~Regev for his help in the data analysis.  
We thank K.~Nollett and A.~Cs\'{o}t\'{o} for a fruitful 
correspondence.  This work was partly supported by the Israel-Germany 
MINERVA foundation and by the Israel Science foundation.

\end{document}